\documentclass[letterpaper,12pt]{article}
\pdfoutput=1
\usepackage{jheppub}
\usepackage{epsfig}
\usepackage{amsmath}
\usepackage{subfigure}
\usepackage{float}

\DeclareSymbolFont{AMSa}{U}{msa}{m}{n}
\DeclareSymbolFont{AMSb}{U}{msb}{m}{n}
\let\Box\relax
\DeclareMathSymbol{\Box}{\mathord}{AMSa}{"03}

\newcommand{\be}{\begin{equation}}
\newcommand{\ee}{\end{equation}}
\newcommand{\bea}{\begin{eqnarray}}
\newcommand{\eea}{\end{eqnarray}}

\newcommand{\f}{\frac}
\newcommand{\h}{\hspace*{1mm}}

\title{Decoupled sectors and Wolf-Rayet galaxies}

\author[1]{Willy Fischler}
\author[2]{, Jimmy}
\author[1]{and Dustin Lorshbough}
\affiliation[1]{Department of Physics and Texas Cosmology Center\\The University of Texas at Austin,
TX 78712.}
\affiliation[2]{Department of Physics and Astronomy\\Texas A\&M University, College Station, TX 77843.}
\emailAdd{Fischler@physics.utexas.edu}
\emailAdd{Jimmy@physics.tamu.edu}
\emailAdd{Lorsh@utexas.edu}

\abstract{It has recently been proposed that gamma-ray burst (GRB) events may be modified by the presence of a dark matter sector subcomponent that is charged under an unbroken U(1). This proposal depends upon there being a non-trivial density of charged dark matter in star forming regions of galaxies which host GRBs.  We discuss four Wolf-Rayet galaxies (NGC 1614, NGC 3367, NGC 4216 and NGC 5430) which should contain comparable amounts of dark matter gas and visible matter gas in the star forming regions.  We show that the ratio of dark jet power to visible jet power depends only on the ratio of particle mass and charge when the densities are equal, allowing for these input parameters to be probed directly by future observations of GRBs.}

\preprint{$\begin{array}{c}\text{UTTG-32-14}\\\text{TCC-029-14}\end{array}$}

\begin{document}
\maketitle
\flushbottom
\section{Introduction}
Recently it has been proposed that dark matter may be charged under a dark unbroken U(1) and decoupled from the Standard Model \cite{Ackerman:mha,Feng:2009mn}.  Ackerman, Buckley, Carroll and Kamionkowski \cite{Ackerman:mha} found that for a dark matter model which includes only a single charged fermion and a dark photon, one cannot simultaneously satisfy bounds arising from galactic dynamics and attain the correct relic density.  It was noted that in order to make this model viable, additional interactions must be added and the case of a fermion additionally charged under SU(2)$_L$ was studied in detail.  Furthermore an explicit model of right-handed stau dark matter based upon the MSSM, with only a single fermion generation, was found to be a viable model by Feng, Kaplinghat, Tu and Yu \cite{Feng:2009mn}.

The presence of an unbroken U(1) allows for the possibility of dark atomic bound states as alluded to in \cite{Ackerman:mha,Feng:2009mn} and more fully explored in \cite{Kaplan:2009de,Kaplan:2011yj,Cline:2012is}.  Dark matter analogs of electrons and protons can interact with dark photons to form hydrogen-like dark atoms as investigated by Kaplan, Krnjaic, Rehermann and Wells \cite{Kaplan:2009de}.  They found a reasonable UV completion of such a model by allowing the decay of heavy right-handed neutrinos to induce both dark sector and visible sector lepton number densities \cite{Kaplan:2011yj}.  These studies introduced a massive gauge boson in order to mediate between the dark sector and visible sector allowing for possible direct detection signatures, though as shown by Cline, Liu and Xue the kinetic coupling between dark and visible photons already leads to interesting physics without introducing additional dark gauge bosons \cite{Cline:2012is}.  For example, in \cite{Foot:2014uba} a small kinetic coupling between dark photons and visible photons was utilized by Foot and Vagnozzi to describe a dynamical halo model which was shown to explain observed galactic structure and result in a daily modulation for direct detection experiments \cite{Foot:2014osa}.

Cyr-Racine and Sigurdson have performed a detailed study of the cosmological and astrophysical implications of atomic dark matter \cite{CyrRacine:2012fz}.  In particular, they found that dark acoustic oscillations can become imprinted on the matter power spectrum providing a strong constraint on the model.  For the case in which all of dark matter is composed of atoms, the constraint that dark matter halos contain essentially collisionless dark matter provides the strongest constraint.  However, if only a fraction of dark matter were composed of atomic bound states this constraint could be relaxed since there are less atoms within any given halo to self interact.

Fan, Katz, Randall and Reece exploited the weakening of constraints by allowing only a fraction of all dark matter to be charged under the dark U(1) \cite{Fan:2013yva,Fan:2013tia}, as alluded to in \cite{CyrRacine:2012fz}.  This model was termed ``partially interacting dark matter" (PIDM).  It was shown that there exists parameter ranges in which the dark atoms may radiatively cool through the emission of dark photons and collapse to form structure such as dark galactic disks.  However, their parameter space studied was shown to be very strongly constrained by observations of the matter power spectrum by Cyr-Racine, de Putter, Raccanelli and Sigurdson \cite{Cyr-Racine:2013fsa} in an analysis similar to \cite{CyrRacine:2012fz}.  The matter power spectrum constraints generically allow PIDM to be at most 4\%-5\% of the dark matter in the universe.

Fan, Katz and Shelton studied possible direct and indirect detection signals associated with PIDM \cite{Fan:2013bea}.  However, it would be preferrable to have a method for probing PIDM which does not rely on the mediator between the dark sector and visible sector.  One proposed method for probing this model of dark matter is to look for the modification to gamma-ray burst (GRB) physics \cite{Banks:2014rsa,Fischler:2014jda}.

Banks, Fischler, Lorshbough and Tangarife noted that it is possible for astrophysical jets of dark radiation to be launched by astrophysical objects \cite{Banks:2014rsa}.  In the Blandford-Znajek mechansim, black hole spin down powers the emission of astrophysical jets.  These jets of dark radiation would therefore extract angular momentum from the black hole.  In this situation there would be a mismatch between the observed power decrease of the visible jet and the spin down of the black hole.  In order for the spin down of the black hole to be appreciable, the density of PIDM and visible plasma must be comparable near the black hole.  The goal of this paper is to identify candidate galaxies in which the density of PIDM and the density of visible matter gas may be comparable in star forming regions which could host GRBs.

In order to discuss the amount of PIDM in star forming regions, one must know about the spatial distribution of PIDM on galactic scales.\footnote{Note, our discussion on PIDM distribution actually applies to any dark matter sector gas.}  This question is very difficult to answer since the ways in which galaxies form and interact is diverse with multiple relevant timescales.  However, we will be able make a statement about how PIDM is distributed throughout galactic scales for at least one class of galaxies by exploiting previous studies of minor merger gas dynamics.

A major merger event occurs between galaxies of comparable masses, the most conservative definition being a mass ratio of 1:1 or 1:2 \cite{Jimmy:2013rta}. Minor merger events, which will be the focus of this paper, occur when a merger takes place between two galaxies which are of significantly different mass.  A defining characteristic of these events is that the properties of the larger galaxy largely dictate the dynamics of the merger.  As we will see in section \ref{sec:gas}, this will allow us to determine the relative amount of PIDM and visible sector gas in star forming regions.

We review some properties and phenomenology of Wolf-Rayet stars and galaxies in section \ref{sec:WolfRayet} .  We will discuss four criteria for galaxies to have a non-trivial density of PIDM in star forming regions which could host GRBs.  We present four candidate galaxies and three possible candidate galaxies which satisfy our criteria.  In section \ref{sec:gas} we discuss how minor mergers allow for there to be a comparable density of PIDM and visible sector gas in star forming regions.  We will discuss the implications of equal density dark and visible plasma in star forming regions for GRB physics in section \ref{sec:GRB}.  We conclude in section \ref{sec:conclusion}.

\section{Wolf-Rayet Stars and Galaxies}\label{sec:WolfRayet}
There are four criteria to consider when searching for galaxies that are likely to have PIDM near GRBs:\\
\hspace*{8mm}1. They are rotationally supported discy galaxies which means their kinematics are well understood and their dark matter profiles could reasonably be measured.\\
\hspace*{8mm}2. They have recently undergone minor mergers with a dwarf galaxy. We will discuss in section \ref{sec:gas} that this means the PIDM and visible sector gas are accreted into the larger galaxy in a known manner.\\
\hspace*{8mm}3. The satellite dwarf galaxy that merges with the massive spiral galaxy is dark matter rich.  For example, observational studies of Blue Compact Dwarf galaxies have shown their dark matter content to be within the range of 80-95\% \cite{Carignan89,Meurer:1998gi,Bureau:2001yf,Elson:2013tz}.\\
\hspace*{8mm}4. They contain Wolf-Rayet stars, which may play a critical role in GRB physics \cite{Kumar:2014upa} as progenitor stars.\\
We will first review the general properties of Wolf-Rayet stars and Wolf-Rayet galaxies before proceeding to discuss four candidate galaxies which satisfy the above criteria.

A typical Wolf-Rayet star is in the mass range 10-25 M$_{\text{Sun}}$ or greater \cite{Crowther:2006dd}.  They begin their life as an O-type star with a mass of at least 25 M$_{\text{Sun}}$ and have short lifetimes of $\sim$5 Myr.  This mass limit assumes a solar metalicity.  As the metalicity decreases, the lower limit on stellar mass increases.  Only $\sim$10\% of the star's lifetime is spent in the Wolf-Rayet phase \cite{Meynet:2004np} making them remarkably short-lived astronomically.  These massive stars produce heavy elements at a high rate as well as dense and powerful stellar winds \cite{Crowther:2006dd}.  As a result, they contribute significantly to the chemical enrichment and feedback of their host galaxy. They are also the primary progenitor candidates for some long GRBs \cite{Kumar:2014upa}.

Wolf-Rayet galaxies are galaxies which contain Wolf-Rayet stars.  We are ultimately interested in how PIDM is distributed with respect to the visible sector gas throughout a Wolf-Rayet galaxy.  To that effect we searched through the literature to identify galaxies from the Wolf-Rayet catalog \cite{Schaerer:1998vt} and found the following candidate galaxies: NGC 1614 \cite{Konig:2013uja}, NGC 3367 \cite{HernandezToledo:2011fj}, NGC 4216 \cite{Paudel:2013bxa} and NGC 5430 \cite{O'Halloran:2005fg}.  Dark matter mass estimates can vary widely from galaxy to galaxy, even within populations such as late-type spirals, however the generally accepted typical value for a spiral galaxy is 10\% visible matter and 90\% dark matter based on observations of stellar rotation curves \cite{Rubin93}.

\subsection{Other Possible Candidates}
Throughout our search of the literature, we found the following galaxies which were close to our criteria, but we were unable to include for the reasons described below.\\
\hspace*{8mm}1. NGC 3003: Although there is reasonable evidence for a merger, the inclusion of this galaxy in the original Wolf-Rayet catalog may have been a mistake \cite{Wilcots01}.\\
\hspace*{8mm}2. NGC 6764: The evidence for a merger within this galaxy is inconclusive \cite{M:2013xva}.\\
\hspace*{8mm}3. NGC 7714: Although it is quite clearly undergoing a very visually stunning merger, by assuming a constant M/L ratio between the two progenitors, it would appear that the mass ratio is on the order of 1:3 so the dynamics are likely to be close to that of a major merger \cite{Peterson:2009cx}.

\subsection{Observational Considerations}
Despite the fact that our selection criteria is not incredibly stringent, we find the small number of galaxies (4-7) found to be quite reasonable. By the Galaxy Mass Function \cite{Tomczak:2013bxa} we would expect dwarf galaxies to be more common than larger galaxies and we would expect a large number of minor mergers. However, several factors conspire to make ideal galaxies for our analysis difficult to find. Firstly, the evidence for a minor merger is difficult to detect due to the small mass influence on the larger galaxy, also evidence for a minor merger disappears faster than evidence for a major merger. Similarly, the short lifetime of Wolf-Rayet stars in the Wolf-Rayet phase means that they are only detectable for a very short timescales ($\sim$0.5 Myrs \cite{Meynet:2004np}).  Therefore it is likely that there are many more galaxies satisfying our selection criteria that we are unable to presently detect.

\section{Visible and Dark Gas Inflow}\label{sec:gas}
Hernquist and Mihos performed numerical simulations of minor merger events between a primary galaxy and satellite galaxy \cite{Hernquist94,Hernquist:1995um}.  In their simulations the halo is represented by a truncated isothermal sphere density profile and the accretion disk profile is taken to be vertically isothermal.  The structure of the primary galaxy bulge is given by the potential-density pair proposed by Hernquist \cite{Hernquist:1990be}.  The gas inflow strength during a minor merger between a rotationally supported massive galaxy and a satellite dwarf galaxy is mainly determined by the structure of the more massive galaxy (in particular, the parameterization of the compact bulge) and is relatively insensitive to the microphysics describing the gas.  It was found \cite{Hernquist:1995um} that approximately 45\% of the visible matter within the primary galaxy disk collapses to a region a few hundred parsecs across.  This high visible gas density was interpreted as capable of producing intense starburst activity.

We will now compare the density of visible sector gas and dark sector gas in star forming regions.  We begin by estimating the amount of visible matter near the center of the galaxy after the completion of the minor merger.  The ratio of visible matter in the galaxy to the total mass of dark matter in the galaxy is given by \cite{Rubin93}
\begin{equation}
\epsilon=\f{M_{\text{gal,vis}}}{M_{\text{gal,DM}}}\approx\f{0.1M_{\text{primary}}}{0.9M_{\text{primary}}}\approx11\%.
\end{equation}
Since the ratio of visible matter comprising a galactic disk to the total mass of dark matter is typically one third of this ratio \cite{Fan:2013yva,Fan:2013tia}, we have
\begin{equation}
\epsilon_{\text{disk}}=\f{M_{\text{disk,vis}}}{M_{\text{gal,DM}}}\approx\f{1}{3}\epsilon\approx3.7\%.
\end{equation}
As previously discussed, 45\% of the visible matter in the disk comprises the star forming region leading to a mass estimate of
\begin{equation}\label{eq:region_vis}
M_{\text{region,vis}}\approx(45\%)M_{\text{disk,vis}}\approx(1.5\%)M_{\text{primary}}.
\end{equation}
Depending on how much visible matter initially comprises the satellite galaxy ($5\%-20\%$ \cite{Carignan89,Meurer:1998gi,Bureau:2001yf,Elson:2013tz}) and how much of the visible matter in the satellite galaxy is able to sink to the star forming region ($\lesssim40\%$ \cite{Hernquist:1995um}), this value may be as large as $\sim3.5\%$.

The dark matter sector gas can be assumed to fall directly into star forming regions since it does not strongly interact with the visible matter nor the dominant component of dark matter.  If we assume that the ratio of PIDM to total dark matter in the satellite and primary galaxies is the same as the cosmological ratio of PIDM to total dark matter, $\sim5\%$, then the mass of PIDM that falls into the central star forming region is simply given by
\begin{equation}
M_{\text{region,dark}}\approx(5\%)(M_{\text{DM,dwarf}}+M_{\text{DM,primary}}).
\end{equation}
Given that the satellite dwarf galaxy mass is 80\%-95\% dark matter \cite{Carignan89,Meurer:1998gi,Bureau:2001yf,Elson:2013tz}, we find that $\sim4\%-4.75\%$ of satellite dwarf galaxy mass will be dark matter sector gas.  Likewise, for the spiral galaxy we initially expect $\sim4.5\%$ of the mass to be dark matter sector gas.  For a 1:n minor merger ($n\gtrsim4$), this means that $\gtrsim4.5\%$ of the initial primary galaxy mass will be in the form of PIDM after the minor merger is complete,
\begin{equation}\label{eq:region_dark}
M_{\text{region,dark}}\approx\left(4.5\%+\f{4\%}{n}\right)M_{\text{primary}}\left(\text{for a 1:n merger}\right).
\end{equation}

While the amount of dark matter sector gas in the star forming region (\ref{eq:region_dark}) can be comparable to the visible sector gas (\ref{eq:region_vis}), a simulation capable of studying the detailed structure of the starburst region is necessary to know how the density ratio around any one given GRB compares.  We leave a detailed numerical analysis of this problem for future work.  We instead will discuss the implications of allowing for a comparable density of dark sector and visible sector gas in star forming regions for GRB studies in the next section.

\section{Implications for Gamma Ray Bursts}\label{sec:GRB}
We now will calculate how the ratio of jet luminosities is affected by the presence of PIDM when dark and visible densities are comparable.  The power emitted from the Blandford-Znajek mechanism depends on the black hole angular velocity, $\Omega_h$, and the magnetic flux threading the black hole, $\Psi$, as \cite{Blandford:1977ds,Komissarov01}
\begin{equation}
\dot{E}_{BZ}\propto \Omega_h^2\Psi^2.
\end{equation}
The only difference between the visible sector and the dark matter sector will be the magnetic flux threading the black hole.  For the case in which a well defined accretion disk forms and feeds the black hole, the flux may be written in terms of the plasma sound speed, $c_s$, and the density, $\rho$, as \cite{Pringle96}
\begin{equation}
\Psi^2\sim (\text{Area})^2B_z^2,\h\h\h B_z^2\approx\left(\f{(\text{disk height})}{(\text{disk radius})}(c_s)\right)^2\rho.
\end{equation}

We note that the plasma sound speed is in general given by $c_s\sim\sqrt{T_e/m_p}$, where $T_e$ is the electron temperature in the plasma and $m_{p/e}$ will denote the mass of the proton/electron respectively.  If we assume that the accretion disk geometry is comparable between the dark plasma and the visible plasma due to the similar gravitational environment, we find that the power ratio is given as
\begin{equation}\label{eq:ratio_jet_temp}
\f{\dot{E}_{\text{BZ,dark}}}{\dot{E}_{\text{BZ,vis}}}\sim\f{T_{e,\text{dark}}}{T_{e,\text{vis}}}\f{m_{p,\text{vis}}}{m_{p,\text{dark}}}\f{\rho_{\text{dark}}}{\rho_{\text{vis}}}.
\end{equation}
The temperature ratio scales as the ratio of accretion luminosity to the (1/4) power \cite{AccretionPower}.  Since the gravitational environment is the same, we will assume that the accretion luminosity for the dark sector and visible sector are some similar constant times their respective eddington luminosity
\begin{equation}
\f{T_{e,\text{dark}}}{T_{e,\text{vis}}}\sim \left(\f{L_{\text{acc,dark}}}{L_{\text{acc,vis}}}\right)^{1/4}\sim\left(\f{L_{\text{edd,dark}}}{L_{\text{edd,vis}}}\right)^{1/4}\sim \left(\f{m_{p,\text{dark}}}{m_{p,\text{vis}}}\f{\sigma_{\text{T,vis}}}{\sigma_{\text{T,dark}}}\right)^{1/4}.
\end{equation}
The Thompson scattering crossing section depends on the fine structure constant and the electron mass allowing us to rewrite this as
\begin{equation}
\f{T_{e,\text{dark}}}{T_{e,\text{vis}}}\sim \left(\f{m_{p,\text{dark}}}{m_{p,\text{vis}}}\right)^{1/4}\left(\f{\alpha_{\text{vis}}}{\alpha_{\text{dark}}}\f{m_{e,\text{dark}}}{m_{e,\text{vis}}}\right)^{1/2}.
\end{equation}
Upon substitution into our power ratio expression (\ref{eq:ratio_jet_temp}), the only remaining dependence is in terms of mass, charge and density
\begin{equation}\label{eq:final}
\f{\dot{E}_{\text{BZ,dark}}}{\dot{E}_{\text{BZ,vis}}}\sim\left(\f{m_{p,\text{vis}}}{m_{p,\text{dark}}}\right)^{3/4}\left(\f{\alpha_{\text{vis}}}{\alpha_{\text{dark}}}\f{m_{e,\text{dark}}}{m_{e,\text{vis}}}\right)^{1/2}\f{\rho_{\text{dark}}}{\rho_{\text{vis}}}.
\end{equation}
Therefore for environments in which the densities are comparable, $\rho_{\text{dark}}\sim\rho_{\text{vis}}$, the ratio of power emitted depends only on the model input parameters of mass and charge.

\section{Conclusions}\label{sec:conclusion}
Astrophysics provides a laboratory to test theories of self-interacting dark matter even if it couples only very weakly to the Standard Model particles.  In order to discover new opportunities for using astrophysics to study the nature of dark matter self-interactions, we have provided criteria for identifying galaxies which should contain a non-trivial amount of self-interacting dark matter in the vicinity of star forming regions.

We have identified four candidate galaxies satisfying these criteria and three additional galaxies that may also satisfy these criteria but require further observations.  The four candidate galaxies are NGC 1614, NGC 3367, NGC 4216 and NGC 5430.  The three possible candidate galaxies are NGC 3003, NGC 6764 and NGC 7714.

In galaxies where the density of dark plasma and visible plasma are comparable near GRBs, the ratio of jet power emitted reduces to an expression in terms of mass and charge ratios for the visible and dark matter plasma constituents.  Therefore future studies of GRBs, as described in \cite{Banks:2014rsa,Fischler:2014jda}, may provide a method of probing the PIDM mass and charge input parameters.

\section*{Acknowledgments}
The work of W.F. and D.L. was supported by the National Science Foundation under Grant Number PHY-1316033.
%

\end{document}